# Network Science approach to Modelling Emergence and Topological Robustness of Supply Networks: A Review and Perspective

Supun S. Perera [1,*]
(E-mail: s.perera@econ.usyd.edu.au Telephone: +61291141893, Fax: +61291141863)

Michael G.H. Bell [1]
(E-mail: michael.bell@sydney.edu.au Telephone: +61291141816, Fax: +61291141863)

Michiel C.J. Bliemer [1]
(E-mail: michiel.bliemer@sydney.edu.au Telephone: +61291141840, Fax: +61291141863)

[1] Institute of Transport and Logistics Studies, University of Sydney, Australia

[*] Corresponding author:
University of Sydney
University of Sydney Business School
Institute of Transport and Logistics Studies (ITLS)
2008 Sydney, NSW
Australia

**Abstract**
Due to the increasingly complex and interconnected nature of global supply chain networks (SCNs), a recent strand of research has applied network science methods to model SCN growth and subsequently analyse various topological features, such as robustness. This paper provides: (1) a comprehensive review of the methodologies adopted in literature for modelling the topology and robustness of SCNs; (2) a summary of topological features of the real world SCNs, as reported in various data driven studies; and (3) a discussion on the limitations of existing network growth models to realistically represent the observed topological characteristics of SCNs. Finally, a novel perspective is proposed to mimic the SCN topologies reported in empirical studies, through fitness based generative network models.

## 1. Introduction

Global supply chain networks (SCNs) play a vital role in fuelling international trade, freight transport by all modes, and economic growth. Due to the interconnectedness of global businesses, which are no longer isolated by industry or geography, disruptions to infrastructure networks caused by natural disasters, acts of war and terrorism, and even labour disputes are becoming increasingly complex in nature and global in consequences (Manuj and Mentzer, 2008). Disruptions ripple through global SCNs, potentially magnifying the original damage.

Even relatively minor disturbances, such as labour disputes, ground traffic congestion or air traffic delays can result in severe disruptions to local and international trade. Therefore, this 'fragility of interdependence' creates new risks to global and local economies (Vespignani, 2010).

At the local level, disturbances to SCNs can have major social and economic ramifications. For instance, during the 2011 Queensland floods in Australia, the key transportation routes were shut down, preventing supermarkets from restocking and leading to critical food shortages (Bartos, 2012). However, at the global level, these consequences can be magnified, resulting in more significant and longer lasting damage. A recent example of such a global SCN disruption is the 2011 Tohoku earthquake and ensuing tsunami in the northeast coast of Japan. Alongside the appalling humanitarian impact, this tsunami caused destruction of critical infrastructure in Japan, resulting in a domino effect, which propagated through global SCNs, with significant global economic consequences. It is reported that for several weeks following the disaster, Toyota in North America experienced shortages of over 150 parts, leading to curtailed operations at only 30% of capacity (Canis, 2011). Similar impacts were observed following the September 11[th] terrorist attacks on the United States in 2001, where movement of electronic and automotive parts were disrupted due to the shutdown of air and truck transportation networks (Sheffi, 2001). These high impact low probability disruptions have affected a large number of economic variables such as industrial production, international trade and logistics operations, thus revealing vulnerabilities in the global SCNs, which are traditionally left unaddressed (Tett, 2011). Therefore, the design of supply chains that can maintain their function in the face of perturbations, both expected and unexpected, is a key goal of contemporary supply chain management (Lee, 2004).

Until recently, the primary focus of supply chain management was on increasing efficiency and reliability by means of globalization, specialization and lean supply chain procedures. Although, these practices enable cost savings in daily operations, they have also made the SCNs more vulnerable to disruptions (World Economic Forum, 2013). Under a low probability high impact disruption, lean supply chains would shut down in a matter of hours, with global implications. Supply concentration and IT reliance make the supply chains vulnerable to targeted attacks. This is particularly evident in the SCNs with low levels of 'buffer' inventory (Jüttner et al., 2003).

A recent strand of publications, by both academic and industry communities, has revealed the importance of understanding and quantifying robustness in global SCNs. Increasing focus has been given to modelling SCNs as complex adaptive systems, in recent years, using network science methods to examine the robustness of various network topologies (Choi et al, 2001; Surana et al., 2005; Brintrup et al., 2016).

The aim of this paper is to present a critical assessment of the research published, mainly in the last decade, in the field of modelling the topology and robustness of SCNs using network science concepts. A novel perspective is then presented in relation to the way forward. The subsequent sections of this paper are structured as follows; Section 2 discusses the complex system nature of modern SCNs and introduces key network science concepts in the context of

SCNs; Section 3 presents the network science approach to modelling the topology and robustness of SCNs; Section 4 presents a discussion, including comparisons, critiques and potential methodological improvements, of the research reviewed, and Section 5 provides conclusions and outlines possible directions for future research.

## 2. From complex systems theory to network science

### 2.1 Complex systems theory

Complex systems theory is a field of science that is used to investigate how the individual components and their relationships give rise to the collective behaviour of a given system (Ladyman et al., 2013). In essence, complex systems possess collective properties that cannot easily be derived from their individual constituents. For example, social systems which comprise relationships between individuals, the nervous system which functions through individual neurones and connections, and life on Earth itself, can all be regarded as complex systems (Kasthurirathna, 2015).

Although complex systems do not have a formal definition, the following three key features broadly characterise such systems (Bar-Yam, 2002);

1. *Emergence*: Macro level properties, which dynamically originate from the activities and behaviours of the individual agents of the system, cannot be easily explained at the agent level alone (Kaisler and Madey, 2009). Therefore, emergence is governed by micro level interactions that are 'bottom-up' rather than 'top-down' rules.
2. *Interdependence*: Individual components depend on each other to varying degrees (Buckley, 2008).
3. *Self-organisation*: This is the attribute that is most commonly shared by all complex systems, where large scale organisation manifests itself spontaneously without any central control, based on local feedback mechanisms that either amplify or dampen disturbances (Mina et al., 2006).

### 2.2 Complex system characteristics of modern supply chain networks

Traditionally, a focal firm is assumed to be responsible for shaping the structure of a given SCN by selecting different suppliers for various purposes, such as reduced cost, increased flexibility/redundancy, and so on. However, the ability of a single firm to shape its supply chain seems to significantly diminish as SCNs become more global and complex in nature. Therefore, the topological structure of a SCN can increasingly be considered as emergent. As such, in a global and a complex business landscape, an individual firm may benefit more from positioning itself within the SCN rather than attempting to shape the SCN's overall topology (Xuan et al., 2011).

Choi et al. (2001) note the complex adaptive system nature of large scale SCNs, where an interconnected network of multiple entities exhibit adaptiveness in response to changes in both the environment and the system itself. System behaviour emerges as a result of the large number of activities made in parallel by interacting entities (Pathak et al., 2007). Therefore, from the point of view of a single firm, the overall SCN is a self-organising system, which consists of various entities engaging in localised decision-making. Given this distributed nature of decision making, the configuration of the final SCN is beyond the realm of control of one organisation. Indeed, individual firms may pursue their own goals with the SCN emerging over time (Choi and Hartley, 1996; Choi et al., 2001).

## 2.3 Network modelling of supply chains

Traditionally, supply chains have been modelled as multi agent (or agent based) systems, in order to represent explicit communications between the various entities involved (Gjerdrum et al., 2001; Julka et al., 2002; Nair and Vidal, 2011). The earliest example of such a model is Forrester's supply chain model (Forrester, 1961; Forrester, 1973), which comprised four types of agents, representing various organisations involved in a supply chain (namely; retailers, wholesalers, distributors and manufacturers), interacting with each other. Such agent-based models (ABMs) provide autonomy to each entity involved and define behaviours in terms of observables accessible to each agent and its goals, norms and decision rules (Parunak et al., 1998; Rahmandad and Sterman, 2008). ABMs are a form of logical deduction, since, given a set of basic rules and initial conditions, the emergent outcomes are embedded in the rules, however surprising they may be (Epstein and Axtell, 1996; Berryman and Angus, 2010).

ABMs are considered to be micro-models, since they facilitate system level inference from explicitly programmed, micro-level rules in simulated agent populations over time and space in a given environment. While such a bottom-up approach maybe suitable for relatively small systems, the exponential increase in the number of connected entities that comprise modern global SCNs favour a top-down approach to system modelling (Pruteanu, 2013).

In this regard, the macro perspective offered by network models are particularly valuable. A recent surge in interest in the area of networks has paved the way for what is now known as 'network science'. Starting from the mathematical field of graph theory (Bondy and Murty, 1978; West, 2001), network science has now matured into a separate field, borrowing concepts from other domains such as statistical mechanics (Albert and Barabasi, 2002; Newman et al., 2011).

Network models focus on how topological properties affect various system properties. Such models typically do not have an environment or coevolution of the environment with the system. Rather, they consist of an ensemble of nodes that behave coherently. This top-down approach considers the network as a single entity and in some models, the individual nodes may exchange information and update the state of the system based on global specifications, which makes the system less prone to unpredictable emergent behaviour (Pruteanu, 2013).

## 2.4 Basic network models

The series of papers published by Erdös and Rényi on random graphs, between 1950 and 1960, sparked initial interest in network science. However, since the introduction of small-world networks by Watts and Strogatz in 1998, interest in the field of network science has surged, as evident in literature.

The following networks are now widely regarded as benchmarks;

1) *Erdös-Rényi (ER) Random graphs:*
   - Nodes are randomly connected to each other.
   - Modelled using the Erdös-Rényi model (Erdös and Rényi, 1959).
2) *Small-world networks:*
   - Most nodes are not neighbours of one another, but can be reached from every other node by a small number of steps.
   - Modelled using the Watts–Strogatz model (Watts and Strogatz, 1998).
3) *Scale-free networks:*
   - The degree distribution follows power-law, at least asymptotically.
   - Modelled using the Barabasi-Albert (BA) model (Barabási and Albert, 1999).

The key characteristics of the above mentioned network topologies are presented in Figure 1.

*2.4.1 Modelling network topology*

A wide range of network models are available in network science literature. They can be broadly categorised into two distinct classes as follows;

1. Generative models – the aim of these models is to generate a snapshot of a topology. Among generative models, some are static (time independent topologies) while some use growth (and other mechanisms). Furthermore, some generative models include predefined global properties (such as degree distribution, hierarchy and modularity) while others predefine a local property (such as the attachment probability).
2. Evolving models – the aim of these models is to capture the microscopic mechanisms underlying the temporal evolution of a network topology. These models include growth and in some instances may include node deletion and link rewiring. For a comprehensive summary of mechanisms underlying various evolving network models, readers are referred to Albert and Barabasi (2002).

Figure 2 outlines the different perspectives offered by generative and evolving network models.

Based on the above classification, both ER random and small-world network models are static generative models as they imply a fixed number of nodes where links are placed between nodes using some random algorithm. These models are therefore less widely used to model dynamical open systems, such as SCNs. However, the ER random model is generally used by researchers as a null model to test whether a real network property is statistically significant or simply attributable to random connectivity (Kito et al., 2014).

It is noted that the result of any static generative model can also be obtained by an evolving model (for example, an ER random network can be conveniently generated using an evolving model with a specified growth process). In fact, any evolving model can be used for generative purposes. In this regard, the BA model, which is generally considered as an evolving network model, can also be used for generative purposes, depending on the study requirement.

An evolving network growth model governs the time evolution of networks by specifying the way in which the new nodes connect with the existing nodes in the network (Zhao et al., 2011(a)). This process is referred to as 'attachment' and various network growth models comprise various 'attachment rules', which subsequently generate networks with distinct topologies as they evolve. For example, the mechanism underlying generation of scale free networks has been successfully captured by the growth (in terms of nodes) and preferential attachment mechanism presented in the BA model (Barabási and Albert, 1999). Under preferential attachment, the probability $p_i$ that a new node makes a connection to an existing node $i$ with degree $k_i$ is given by:

$$p_i = \frac{k_i}{\sum_{j \in N} k_j}$$

where $N$ is the set of nodes to which the new node could connect.

The BA model represents a 'rich get richer' process and the resulting scale-free network topology can be used to model many real world networks, such as the World Wide Web, power grids, metabolic networks and social networks (Surana et al., 2005). This concept explains the existence of 'hubs' (a few nodes with a large number of connections), which is a defining feature of scale-free networks.

The degree distribution $P_k$ of a scale free network is approximated with power-law as follows;

$$P_k \sim k^{-\gamma}$$

where $k$ is the degree of the node and $\gamma$ is the power-law exponent.

Many network properties depend on the value of the power-law exponent, $\gamma$ (Barabasi, 2014). Therefore, it is important to accurately estimate the power-law exponent of the degree distribution of a given network topology, as this enables us to compare network topologies on a continuous spectrum. Newman (2005) presents a reliable methodology accurately estimating the power-law exponent of a given degree distribution, which involves plotting the complementary cumulative distribution function. This method does not require data binning and as a result eliminates the plateau observed in linear binning approach for high degree regime by extending the scaling region. Interested readers are referred to Clauset et al. (2009), for a comprehensive review of power-law distributions in empirical data.

*2.4.1.1 Fitness based network models*

In the BA model, it is assumed that a node's growth rate (in terms of new link acquisition) is determined solely by its degree. Accordingly, it predicts that the oldest node always has the most links – this concept is often referred to as the first mover advantage in the economics literature. However, this approach does not take into consideration the intrinsic characteristics of the nodes which may influence the rate at which they acquire new links. For instance, in many real world networks such as Hollywood actor networks and global business networks, some nodes despite being latecomers, acquire links within a short timeframe whereas others are present within the network from early on but fail to acquire high numbers of links (Barabasi, 2014). As such, modelling SCN growth based on a growth model which views new link acquisitions from a purely topological perspective may not be suitable.

Therefore, rather than relying entirely on the node degree, the attachment probability and subsequent network growth should rely on a more basic factor, referred to as node 'fitness' (Caldarelli et al., 2002; Ghadge et al., 2010; Smolyarenko, 2014). The concept of 'fitness' can be thought of as the amalgamation of all the attributes of a given node that contribute to its propensity to attract links, which could also include the node degree.

In order to overcome the limitations mentioned above, a model was proposed by Bianconi and Barabasi (2001). This model is referred to as the Bianconi-Barabasi Model (hereinafter referred to as the BB model) and has the following characteristics;

- *Growth* – At each time step, a new node $j$ with $m$ links and fitness $\phi_j$ is added to the network. In generating an ensemble of networks, $\phi_j$ is sampled from a fitness distribution. Once assigned, the fitness of a node remains constant.
- *Preferential Attachment* – the probability of a new node connecting to node $i$ is proportional to the product of node $i$'s degree $k_i$ and its fitness $f_i$;

$$p_i = \frac{k_i f_i}{\sum_{j \in N} k_j f_j}$$

As can be seen from the above formulation, between two nodes $i$ and $j$ with the same fitness $(f_i = f_j)$, the one with the higher degree will have the higher probability of selection. Conversely, between two nodes $i$ and $j$ with the same degree $(k_i = k_j)$ the node with the higher fitness will be selected with a higher probability, thus indicating that even a relatively new node, with only a few links, can acquire more new links rapidly, if it has a higher level of fitness. As such, consideration is given to both fitness and the degree of the nodes in the above growth model.

More recently, Ghadge et al. (2010) proposed a purely fitness based network growth model, which accounts for the various factors that contribute to the likelihood of a new node being attracted to an existing node within a network. Such fitness based attachment models can indeed be categorised as generative models. This type of models offers greater flexibility owing to their ability to reproduce network topologies with fixed global properties.

In the model proposed by Ghadge et al. (2010), the fitness $f_i$, which represents the propensity of node $i$ to attract links, is formed from the product of relevant attributes $\{f_{i1}, \ldots, f_{i|L|}\}$;

$$f_i = \prod_{k \in L} \varphi_{ik}$$

Where each attribute, $\varphi_i$ is represented as a real non-negative value. Subsequently, it is assumed that the number of attributes affecting a node's attractiveness is sufficiently large and are statistically independent. Therefore, by a version of the Central Limit Theorem, the overall fitness $\phi_i$ will tend to be lognormally distributed, regardless of the type of distribution of the individual factors (Nguyen and Tran, 2012). In a SCN context, the attributes, which contribute to fitness, could include cost, service or product quality, reliability, and so on. Finally, the probability of connecting a new node $j$ to an existing node $i$ is taken to be proportional to its fitness $\phi_i$, as follows;

$$p_i = \frac{f_i}{\sum_{j \in N} f_j}$$

The above attachment rule, named the 'Lognormal Fitness Attachment' (LNFA), differs from the BA model in that node fitness replaces node degree (Nguyen and Tran, 2012). Therefore, in LNFA, a new node which has a large fitness, despite being in the network for a short period of time, can make itself a preferential choice for new nodes entering the network.

Recent work by Bell et al. (2017) have investigated the evolutionary mechanisms that would give rise to a fitness-based attachment process. In particular, it is proven by analytical and numerical methods that in homogeneous networks, the minimisation of maximum exposure to unfitness by each node, leads to attachment probabilities that are proportional to fitness. This result is then extended to heterogeneous networks, with strictly tiered SCNs being used as examples.

*2.4.1.2 Generating null models using configuration model*

Similar to the LNFA model, the configuration model belongs to the wider class of network generative models. A generative model allows us to choose parameters and draw a single instance of a network. Since a single generative model can generate many instances of networks, the model itself corresponds to an ensemble of networks.

The configuration model is commonly used to generate networks with pre-defined degree sequences. It is particularly useful for generation of null models for the purposes of hypothesis testing. Comparison of properties of an empirical network with the properties of an ensemble of networks generated by the configuration model allows one to identify if the properties observed in the empirical network are unique and meaningful or whether they are common to all networks with that degree sequence (Fosdick et al., 2016). When data is available for an empirical network, a technique termed degree preserving randomisation (DPR) is often used in literature to generate random networks which correspond to the configuration model. DPR involves rewiring the original network, to generate an ensemble of null models, while preserving the degree vector (Maslov and Sneppen, 2002). At each time step, the DRP process randomly picks two connected node pairs and switch their link targets. This switching is repeatedly applied to the entire network until each link is rewired at least once. The resulting network represents a null model where each node still has the same degree, yet the paths through the network have been randomised.

For example, Becker et al. (2014) have constructed a manufacturing system network model from real world data (where nodes represent separate work stations and links represent material flows between work stations). By applying the DPR process to generate an ensemble of networks with the same degree distribution, the authors observe that nodes (work stations) with a particularly high betweenness centrality are over-represented in the manufacturing system studied. They concludes that the manufacturing system topology is therefore non-random and favours the existence of a few highly connected work stations.

## 3. Network science approach to modelling the topology and robustness of SCNs

So far, the published research in the area of modelling SCNs as complex networks have demonstrated that a network perspective can indeed be used to successfully represent a supply chain as nodes and links (Thadakamalla et al., 2004; Xuan et al., 2011; Zhao et al., 2011(a); Zhao et al., 2011(b); Wen and Guo, 2012; Li et al., 2013; Li, 2014; Mari et al., 2015 and Kim et al., 2015). A typical SCN model consists of nodes, which represent individual firms (such as suppliers, manufacturers, distributers and retailers), and links, which represent interactions between nodes (such as exchange of information, transportation of material, and financial transactions). Such abstractions can be beneficial in identifying the properties of various types of SCN, as representing too many details could be detrimental to identifying the network properties (Shen et al., 2006). On the other hand, important node or link information could be lost. The level of detail to be represented by a given complex network model is an important decision to be made by the network scientist (Kasthurirathna, 2015).

Given the ultimate goal of obtaining generalizable results for real world SCNs, the theoretical research in this area should be well informed by empirical studies. In particular, empirical studies should be used to establish the key characteristics that need to be represented in the network topologies being generated by a given growth mechanism. Figure 3 illustrates the general methodological framework of research on topological modelling and robustness analysis of SCNs.

### 3.1 Modelling SCN topologies through growth models

In order to characterise the dynamical processes on complex SCNs, the first step is to construct realistic network growth models. Such models can be used to generate an ensemble of networks with the required topological properties, from which insights can be gained into the relationship between the topology and the dynamics of complex networks (Bianconi, 2016).

In the context of supply chains, the concept of growth describes how newcomer firms join existing firms in a SCN. As new entrants join the SCN, trading partners are assigned from within the network. In the above regard, the BA model, despite its simplicity and elegance, includes a number of known limitations, as listed below against their respective implications for SCNs.

*Table 1: Limitations of BA Model in Modelling SCNs*

| Limitation | SCN Modelling Implication |
|---|---|
| Does not account for internal link formations (Barabasi, 2014) | In a SCN, new links may not only arrive with new firms but can be created between the pre-existing firms. |
| Cannot account for node deletion (Barabasi, 2014) | Firms may exit a given SCN over time. |
| An isolated node is unable to acquire any links since according to preferential attachment, the probability of a new node connecting to an isolated node is strictly zero (since the connection probability is governed by the existing number of connections). | In reality, any firm has a certain level of initial attractiveness. |
| Assumes that all firms within the supply network are homogeneous in nature with no differentiation other than the topological aspects (Hearnshaw and Wilson, 2013). | Real SCNs include firms with high levels of heterogeneity beyond the number of dealings or connections with other firms. |
| The key requirement of the preferential attachment rule is that every new node joining the network must possess complete and up-to-date information about the degrees of every existing node in the network. | Such information is unlikely to be readily available in a real world setting – for example, when considering a manufacturer for a new partnership, full information about the number of their current suppliers and clients is unlikely to be available (Smolyarenko, 2014). Therefore, an algorithm which relies on local information is deemed more suitable. For example, see Vázquez (2003). |
| Network growth by preferential attachment produces a decaying clustering coefficient as the network expands. | May not be a realistic representation of exchange relationships and concentration of power in firms within the real SCNs (Hearnshaw and Wilson, 2013). |

Indeed, firm partner selection is, in reality, a multi-objective problem and involves numerous factors, such as price, performance, quality, goodwill, transport cost (Jain et al., 2009; Li et al., 2013). However, it is not practical to consider all these factors, so researchers have adopted simple yet intuitive approaches which extend the basic BA model concept. These include specifying selection (attachment) rules based on basic topological properties, such as the node degree (i.e. the connectedness of a given firm), in conjunction with other parameters (such as the number of links entering the system with each type of new node, the rewiring probability, and the topological distance between source and target nodes). Examples of such customised attachment rules are summarised in Table 2.

*Table 2: Key features of customised attachment rules used in literature*

| Customised Attachment Rule | Key Features |
|---|---|
| Ad hoc attachment rules based on the military supply chain example, used by Thadakamalla et al. (2004) | Three types of nodes can enter the system in a pre-specified ratio. Each type of node has a specific number of links. Attachment rule depends on the type of node entering the system. The first link of a new node entering the system attaches to an existing node preferentially, based on the degree. The subsequent links, entering the system with each new node, attach randomly to a node at a pre-specified topological distance (also referred to as the 'hop count', which denotes the least number of links required to be traversed in order to reach a given node from another). |
| Degree and Locality based Attachment | A node entering the system considers both the degree and the distance of an existing node, when establishing connections. In particular, attachment preference for the first link arriving with each new node is calculated |
| (DLA), used by Zhao et al. (2011a). | preferentially based on the degree of the existing nodes. If the node is allowed to initiate more than one link, the subsequent links will attach preferentially to existing nodes based on topological distance. Tunable parameters are used to control the responsiveness of attachment preference to both the degree and the topological distance. |
| Randomised Local Rewiring (RLR), used by Zhao et al. (2011b) | This model is applied to an existing network, by iterating through all links and considering the nodes at either end of each link. With a predetermined rewiring probability, to control the extent of rewiring, each link will disconnect from the highest degree node it is currently connected with and reconnect with a randomly chosen node within a pre-specified maximum radius (which can either be geographical or topological). |
| Evolving model used by Zhang et al. (2012) and Li and Du (2016) | Start with a random network that consists of a pre-specified number of supply nodes with randomly assigned (x, y) coordinates. Supply and demand nodes are sequentially added to the system, according to a pre-specified supply-demand ratio. If the new node is a supply node, the first link will connect to an existing supply node in the system while other links are connected randomly to existing nodes. If the new node is a demand node, all links will connect with existing supply nodes in the system, with connection probability based on the product of degree and the geographical distance. Similar to DLA discussed above, tunable parameters are used to control the responsiveness of attachment preference to both the degree and the geographical distance. |

Each of the aforementioned attachment rules, over time, generate networks with distinct topologies. It is evident, that when constructing network topologies representative of SCNs, the attachment preference is generally governed by three factors;

1) The type of node entering the system (which determines the number of links that enter the system with addition of new nodes and to which existing nodes these links will be connected).
2) Type and degree of existing nodes (preference is given to existing high degree nodes over the low degree nodes, representing the market power and visibility of highly connected firms).
3) Type and topological or geographical distance of existing nodes (preference is given to closer nodes than farther away ones, in terms of either the topological or the geographical distance, representing the 'relationship distance' or the cost of goods movement, respectively).

### 3.2 Concept of SCN robustness

From the contemporary literature in the area of modelling SCN robustness, it is evident that the terms resilience and robustness have been used interchangeably by researchers. However, in the field of network science, the terms resilience and robustness have distinct meanings. For example, a system is called robust, if it can maintain its basic functions in the presence of internal and external perturbations. Hence, a robust SCN would include redundant or parallel components, which if needed can be relied upon to maintain the overall functionality.

In contrast, resilience is defined as the capability of a system to adapt to internal or external perturbations by changing mode of operation, without losing its ability to function (Barabasi, 2014). Therefore, a resilient supply chain should respond quickly and effectively to a given perturbation, such as a change in supply or demand, or to the failure of a component (such as a firm or a material transport route) within the system. The response mechanism of a resilient SCN is attributable to its flexibility to rewire the lost connections away from disrupted nodes (Sheffi and Rice, 2005). As opposed to resilience, the robustness of a SCN does not relate to response mechanisms – it merely reflects the extent to which a given SCN can withstand loss of its components, without losing its basic functions. It is worth noting that most studies have focussed mainly on the topological robustness of SCNs, rather than their resilience.

*3.2.1  Analytical measurements of SCN topological robustness*

Network science offers a rich set of tools for topological robustness analysis. Refer to Costa et al. (2007) and Rubinov and Sporns (2010) for a comprehensive range of measures used for the characterization of complex networks. Some key metrics used in network science, and their corresponding SCN implications at node and network level are presented in Appendix A and B, respectively.

As can be seen from the metrics presented in Appendix A and B, the analytical measures in network science can be used to gain important insights on network structure and robustness quickly and with low computational difficulty. However, the key limitation of using analytical measures is that they are unable to account for the heterogeneity of nodes, in terms of their functionality within a given SCN, since the metrics consider all nodes to be homogeneous in function. In order to overcome the above limitation, researchers have relied on simulations to analyse the topological robustness of SCNs. Furthermore, simulations allow flexibility in analysis through customised robustness metrics (see section below).

*3.2.2  Using simulations to determine the topological robustness of SCNs*

Node failures in networks can be categorised either as 'random failures' or 'targeted attacks'. Random failures entail the same probability of failure across each node within a given network. In contrast, a 'targeted attack' refers to when an attacker selectively compromises the nodes with probabilities proportional to their degrees, where highly connected nodes are compromised with higher probability (Ruj and Pal, 2014).

In the network science literature, random failures and targeted attacks in networks are typically simulated as follows;
1) *Random failure*: Each robustness metric established for the network is recorded at each time step by randomly removing the nodes from the network.
2) *Targeted attacks*: Each robustness metric established for the network is recorded at each time step by sequentially removing the nodes, based on their degree, removing higher degree nodes first, from the network.

The robustness values recorded for each metric, for each network considered, are then compared. It has, so far, been established that the random networks respond similarly to both random failures and targeted attacks. In comparison, the scale-free networks are robust against random failures but are highly sensitive to targeted attacks (Albert et al., 2000). This is due to the presence of hubs (highly connected nodes) in scale-free networks, which are the nodes targeted by an attacker.

A number of researchers have modelled various SCN topologies under both random failures and targeted attacks, and attempted to establish an optimal topology which can withstand each type of failure without compromising the overall network functionality (Thadakamalla et al., 2004; Zhao et al., 2011a, Li and Du, 2016). Each study has established a set of robustness metrics, in order to assess and compare the robustness of each network topology simulated under random failures and targeted attacks. These robustness metrics are variations of the existing standard topological metrics from network science. The most commonly used network topology metrics in supply network research are;

1. *Size of the largest connected component (LCC) of a network* - As nodes are sequentially removed, the graph disintegrates into sub-graphs. The number of nodes in the LCC (or the largest sub-graph) of a fragmented network therefore provides insights into its overall connectivity.
2. *Average or maximum path length in the LCC* - The average or maximum shortest path length between any two nodes in the largest connected component of a network. This provides insights into the overall accessibility of the network.

The above metrics consider the roles of entities (nodes and links) within a distribution network to be homogeneous. Such an assumption is far from reality, since the entities within a real-world supply network play different roles with different characteristics – for example, the distance between two supply nodes or two demand nodes are not as important as that between a supply and a demand node (Zhao et al., 2011b).

Therefore, various researchers (such as Thadakamalla et al., 2004; Zhao et al., 2011a and Zhao et al., 2011b, Xu et al., 2014) have developed new metrics, which realistically represent the heterogeneous roles of nodes within the network. For example, Zhao et al. (2011a) have developed the following customised robustness metrics for distribution networks;

- Supply availability rate is represented as the percentage of demand nodes that have access to supplies from at least one supply node.
- The network connectivity is determined through the size of the largest functional sub-network (LFSN), namely the number of nodes in the LFSN in which there is a path between any pair of nodes and there exists at least one supply node.
- Accessibility is determined by;
    - Average supply path lengths in the LFSN, i.e. the average shortest supply path length between all pairs of supply and demand nodes in the LFSN.
    - Maximum supply path lengths in the LFSN, i.e. the maximum shortest path length between any pair of supply and demand nodes in the LFSN.

### 3.3 Empirical studies on SCN topologies

A review of contemporary literature on SCN topologies reveals that only a limited number of data driven studies are available in this domain. This is mainly due to difficulty in obtaining specific information about supplier/customer relationships, which is often proprietary and confidential. The following table presents a summary of a number of empirical studies available to date. It is noted that these studies generally focus on overall topological character of SCNs rather than robustness.

*Table 3: Summary of empirical studies of SCN topologies*

| Study | Data Source and SCNs Considered | Key Findings |
|---|---|---|
| Parhi (2008) | Customer-supplier linkage network in the Indian auto component industry has been considered (618 firms), using the data from the Auto Component Manufacturers Association of India. | The Indian auto component industry SCN was found to be scale free in topology, with a power-law exponent, $\gamma=2.52$*. |
| Keqiang et al. (2008) | Guangzhou automotive industry supply chain network has been investigated. Data has been collected from 94 manufacturers, between November 2007 and January 2008. | Guangzhou automotive industry SCN was found to be scale-free in topology. Based on the data presented by the authors, we have calculated the power-law exponent of the degree distribution, $\gamma$ to be 2.02. |
| Kim et al. (2011) | Three case studies of automotive supply networks (namely, Honda Accord, Acura CL/TL, and Daimler Chrysler Grand Cherokee) presented by Choi and Hong (2002). | This study has developed SCN constructs based on a number of key network and node level analysis metrics. In particular, the roles played by central firms, as identified by various network centrality measures, have been outlined in the context of SCNs. |
| Büttner et al. (2013) | Present network analysis results for a pork supply chain of a producer community in Northern Germany. Data has been obtained by the producer community for a period of 3 years. | Reports that the degree distribution of the SCN follows power law (in and out degree distributions follow power-law with power-law exponents, $\gamma=1.50$ and $\gamma=1.00$, respectively). Disassortative mixing** has been observed in terms of node degree. |
| Kito et al. (2014) | A SCN for Toyota has been constructed using the data available within an online database operated by Marklines Automotive Information Platform. | The authors have identified the tier structure of Toyota to be barrel-shaped, in contrast to the previously hypothesized pyramidal structure. Another fundamental observation reported in this study is that Toyota SCN topology was found to be not scale free. |
| Brintrup et al. (2015) | Airbus SCN data obtained from Bloomberg database. | Reports that the Airbus SCN illustrates power-law degree distribution, i.e. scale free topology, with a power-law exponent, $\gamma=2.25$*. Assortative mixing was observed based on node degree and community structures were found based on geographic locations of the firms. |
| Gang et al. (2015) | Authors have investigated the urban SCN of agricultural products in mainland China. Data collection is based on author observations over 2 years. | The SCN of agricultural products was found to be scale free in topology, with a power-law exponent, $\gamma=2.75$. High levels of disassortative mixing** has been observed in terms of node degree. |
| Orenstein (2016) | SCN data for food (General Mills, Kellogg's and Mondelez) and retail (Nike, Lowes and Home Depot) | The SCNs considered in this study were found to have scale free topologies with $\gamma < 2$. In particular, for the food industry SCNs for General Mills, Kellogg's and Mondelez were found to have $\gamma = 1.25, 1.47$ and $1.56$, respectively. For the retail |

|  | industries have been obtained from Bloomberg database. | industry, the SCNs for Nike, Lowes and Home Depot were found to have $\gamma$ = 1.83, 1.73 and 1.67, respectively. |
|---|---|---|
| Perera et al. (2016b) | Analysis has been undertaken for 26 SCNs (which include more than 100 firms) out of 38 multi echelon SCNs presented in Willems (2008) for various manufacturing sector industries. | 22 out of the 26 SCNs analysed display 80% or higher correlation with a power-law fit, with power-law exponent $\gamma$=2.4 (on average). Furthermore, these SCNs were found to be highly modular** and robust against random failures. Also, disassortative mixing** was observed on these SCNs. |
| Sun et al. (2017) | A GIS based SCN structure has been simulated for the automobile industry using the data of top twelve car brands of Chinese market in recent five years as basic parameters. | The Chinese automobile SCN simulated using real world data as basic parameters, indicates that the degree distribution conforms to the power-law, with a power-law exponent, $\gamma$=3.32. |

*Note that in some research papers, the power-law exponent is presented for the cumulative degree distribution. In such cases, the power-law exponent of the degree distribution has been established by adding 1 to the power-law exponent of the cumulative degree distribution since the power-law exponent of the cumulative degree distribution is 1 less than the power-law exponent of the degree distribution (Newman, 2007). These instances have been identified with an asterisk in Table 2.

**Refer to Appendix A for detailed definitions (including mathematical formulations) of these metrics.

## 4. Discussion of literature – limitations and improved methodological directions

This section will critically discuss the existing methodologies in contemporary literature, on modelling SCN topologies and robustness.

### 4.1 Modelling SCN Topologies

#### 4.1.1 Insights revealed by empirical studies

While SCNs in real world may not evolve through a single mechanism, it is possible to infer general growth and design principles from the global properties of existing SCNs. In this regard, empirical studies play a major role in pointing the theoretical research work on modelling SCN topologies in a meaningful direction.

A number of past theoretical studies have relied upon the BA model for SCN growth and/or benchmarking purposes (Thadakamalla et al., 2004; Xuan et al., 2011; Zhao et al., 2011(a)). However, based on the results of empirical studies summarised in this paper, it is understood that the BA model (due to its minimal nature) cannot sufficiently represent the growth mechanism underlying SCNs, due to the following:

1) The BA model generates networks with a constant power-law exponent, $\gamma = 3$, as shown by both analytical and simulation methods in Barabasi (2014). The SCNs reported in empirical studies indicate topologies with $\gamma \approx 2$ (it is noted that $\gamma = 2$ is the boundary between hub and spoke ($\gamma<2$) and scale-free ($\gamma>2$) network topologies).
2) The BA model cannot generate networks with pronounced community structure, which has been observed in real SCNs, since all nodes in the network belong to a single weakly connected component (Newman, 2003).
3) Assortative (or disassortative) mixing as observed in real SCNs, is not a feature of networks generated by the BA model - as shown analytically (in the limit of large network size) by Newman (2002).

As can be seen above, although some real world networks have been convincingly modelled by the preferential attachment mechanism presented in the BA model (Barabasi et al., 1999, Albert et al., 1999), this is not so for SCNs. Therefore, a convincing network growth mechanism for SCNs is yet to be formulated.

#### 4.1.2 Suitability of network models in literature for SCN modelling purposes

Considering the limitations of the BA model in representing the topologies of real SCNs (as discussed in the previous section), the generative models which predefine a global property (such as the degree distribution, hierarchy, modularity, etc.) are a good starting point for the researchers in the SCN domain, especially when the interest of research is directed towards understanding the role of network topology on its robustness. In particular, the SCNs in the real world may have evolved based on various non generalizable principles. Therefore, when aiming to study and understand the topological character of SCNs, researchers will benefit more from simply mimicking the observed topologies, than trying to understand the underlying growth mechanism – which may indeed be complex and non-generalizable, beyond the realm of a single mathematical algorithm.

Generative models allow the researchers to recreate a network topology, as observed at a single cross section in time, and undertake further investigations on various phenomena, such as topological efficiency, robustness etc. When information on the adjacency matrix is available for real-world networks, one can simply use the DPR to generate an ensemble of random networks (which correspond to the configuration model) to establish whether the degree distribution on its own is sufficient to describe the property observed in the network at hand. It is worth noting that in tiered SCNs, the DPR process should be restricted to each tier, in order to ensure that links are not swapped between non-compatible tiers.

In many cases, adjacency matrix information for real networks are not readily available. Such situations require the researcher to recreate the degree distribution of the SCN, using only the basic network metrics (such as the power-law exponent) or simply through qualitative descriptions. In this regard, fitness based generative models have recently gained prominence in theoretical research (Caldarelli et al., 2002; Bedogne et al., 2006; Smolyarenko, 2014; Perera et al., 2016a). In fitness based models, the fitness distribution and the connection rules are given by a priori arbitrary functions, which enables considerable amount of tuning (Smolyarenko, 2014). Indeed, this tunability makes such models a useful and practical modelling tool.

For example, the LNFA includes a tunable parameter $\sigma$ (the shape parameter of the lognormal distribution), which can be manipulated to generate a large spectrum of networks. At one extreme, when $\sigma$ is zero, all nodes have the same fitness and therefore at the time a new node joins the network, it chooses any existing node as a neighbour with equal probability, thus replicating the random graph model. On the other hand, when $\sigma$ is increased beyond a certain threshold, very few nodes will have very large levels of fitness while the overwhelming majority of nodes have extremely low levels of fitness. As a result, the majority of new connections will be made to a few nodes which have high levels of fitness. The resulting network therefore resembles a monopolistic/"winner-take-all" scenario, which can sometimes be observed in the real world (however, in some instances, it may be necessary to place a restriction on the highest degree achievable by a single node, in order to represent the 'contractual capacity' of firms). Between the above two extremes (random and monopolistic) lies a spectrum of power law networks which can closely represent many real networks (Ghadge et al., 2010). Figure 4 illustrates the spectrum of network topologies generated by the LNFA model.

Nguyen and Tran (2012) have illustrated that the LNFA model can indeed generate network topologies with $\gamma \approx 2$, which represents many observed SCN topologies in empirical research work (Büttner et al., 2013; Orenstein, 2016). However, the ability of LNFA to generate modular and disassortative networks, as observed in SCNs, remains an open research question.

*4.1.3   Directionality of links*

The inter-firm relationships in SCNs are generally modelled using undirected links. However, the links between nodes in a SCN can include a direction, depending on the specific type of relationship being modelled. The inter-firm relationships in a SCN can be broadly categorised into three classes, namely; (1) material flows, (2) financial flows, and (3) information exchanges. Material flows are usually unidirectional from suppliers to retailers, while financial flows are unidirectional in the opposite direction. Both material and financial flows mostly occur vertically, across the functional tiers of a SCN (however, in some cases, two firms within the same tier, such as two suppliers, could also exchange material and finances) (Lazzarini et al., 2001). In contrast, information exchanges are bidirectional (i.e. undirected) and includes both vertical and horizontal connections (i.e. between firms across tiers and between firms within the same tier). Therefore, the same SCN can include different topologies based on the specific type of relationship denoted by the links in the model. For instance, unlike material and financial flows, SCN topology for information exchanges can exhibit shorter path lengths and high clustering due to relatively larger number of horizontal connections (Hearnshaw and Wilson, 2013).

Compared to undirected network representation, in directed networks, the adjacency matrix is no longer symmetric. As a result, the degree of a node in a directed network is characterised by both in-degree and out-degree. On this basis, the degree distribution of directed networks is analysed separately for in and out degrees. Also, unlike undirected networks, in directed networks the distance between node $i$ and node $j$ is not necessarily the same as the distance between node $j$ and node $i$. In fact, in directed networks, the presence of a path from node $i$ to node $j$ does not necessarily imply the presence of a path from node $j$ to node $i$ (Barabasi, 2014). This has implications on node centrality metrics, such as closeness and betweenness. In addition, it is noted that many dynamics, such as synchronizability and percolation, are different in directed networks compared to undirected networks (Schwartz et al., 2002; Park and Kim, 2006). Therefore, when modelling SCNs, it is important to first identify the specific type of relationship denoted by the links, so that network can be correctly represented as undirected or directed.

*4.2   Additional considerations for robustness testing*

From a SCN point of view, the position of an individual firm with respect to the others can influence both its strategy and behaviour (Borgatti and Li, 2009). Accordingly, analysis of each firm's role and importance based on its position in the SCN can reveal important properties, such as its structural robustness. In this regard, future studies could simulate targeted attacks based on node centrality measures, such as betweenness and closeness centrality, rather than node degree. Such considerations will capture the critical nodes in various perspectives. Also, as Piraveenan et al. (2012) notes, when simulating targeted attacks on empirical networks, one could also rank nodes on the basis of non-topological attributes (such as firm size, output, and geographic location).

Depending on the structure of the overall SCN, disruptions can be experienced in various forms, such as; supply disruptions, logistics disruptions, coordination disruptions and demand disruptions (Yi et al., 2013). These various disruptions can be attributed to either nodes or links or both, for modelling purposes. So far, the focus of modelling has been on unweighted links in SCNs, which essentially indicate that all relationships are considered to be homogeneous in terms of their relative importance. However, real SCNs exhibit large levels of heterogeneity in the capacity and intensity of the connections (links) between the nodes. Rui and Ban (2012) state that empirical observations have shown the existence of nontrivial correlations and associations between link weights and topological quantities in complex networks. In the context of SCNs, the connections, be they physical flows or relationships between organisations, are heterogeneous in terms of the strength and importance. Therefore, the SCN can be better reflected and understood in terms of weighted networks, where weights reflect volume, frequency or the criticality of flows (Hearnshaw and Wilson, 2013). If such information is available, targeted attacks could also be simulated by link removal on the basis of link weights.

## 5 Conclusions and future directions

This paper has presented a comprehensive and critical review of the research undertaken on the use of network science techniques to model the topology and robustness of SCNs. The key challenge in this research is the tailoring of network science principles to SCNs, by identifying the fundamental SCN features. Although network science offers a rich conceptual representation of SCNs, a number of potential improvements to the existing modelling approach have been identified and are proposed for future research.

From the literature reviewed, it is evident that most of the previous research undertaken in the field of modelling SCNs as complex networks have given primary consideration to network topology. Based on the empirical studies, it is evident that most real world SCNs tend to have power-law exponents which fluctuate around 2. It is noted that $\gamma = 2$ is the boundary between hub and spoke ($\gamma<2$) and scale-free ($\gamma>2$) network topologies. Also, most SCN topologies indicate disassortative mixing and modularity (the presence of communities). The well-known BA model is not able to generate network topologies with the above mentioned features. Therefore, researchers are advised to focus on generative models to mimic the SCN topologies observed in empirical studies. This approach is deemed more effective and reliable than the existing methodology of investigating the mechanisms underlying SCN evolution, particularly since the overarching goal of research in this area is to understand the role of network topology on properties such as robustness. It is emphasised that future theoretical work on development of SCN growth models should ideally aim to reflect the above outlined topological features in the network topologies obtained from generative models. A natural extension of this work would be to investigate the ability of fitness based growth models, coupled with node and link heterogeneity, to mimic the topological features, such as modularity and disassortativity, of real world SCNs.

So far, empirical studies have investigated a cross sectional view of real SCNs at a given point in time. However, databases such as Bloomberg offer rich data sets to investigate the evolution of SCNs across time. Therefore, researchers could investigate the evolution of SCNs, using temporal data. Such empirical tests can validate the theoretical network growth models developed so far in the literature.

# 6 Declarations

## 6.1 Ethical Approval and Consent to participate

Not applicable.

## 6.2 Consent for publication

Not applicable.

## 6.3 Availability of supporting data

Not applicable.

## 6.4 Funding

The authors would like to acknowledge the Australian Research Council (ARC) for funding this work under grant DP140103643.

## 6.5 Authors' information

SP carried out the initial review of literature and prepared the draft manuscript. M.G.H. Bell and M.C.J. Bliemer participated in discussions where limitations in existing research methods were identified and improvements proposed. SP and M.G.H. Bell prepared the final manuscript. All authors read and approved the final manuscript.

# APPENDIX A: LIST OF NETWORK LEVEL METRICS AND THEIR SCN IMPLICATIONS

**TABLE A: Network level metrics and their SCN implications**

| Mathematical representation | SCN Implication |
|---|---|
| **Average degree ($<k>$)** | |
| $$<k> = \frac{\sum_i k_i}{N}$$ where $N$ is the total number of nodes in the network | Indicates, on average, how many connections a given firm has. Higher average degree implies good inter-connectivity among the firms in the SCN, which is favourable in terms of efficient exchange of information and material. <br><br> In directed networks, the in-degree characterises the number of supply channels while the out-degree indicates the number of sales channels, of a given firm. |
| **Network diameter** | |
| $$\text{diameter} = \max_{i,j} l(i,j)$$ where $l$ is the number of hops traversed along the shortest path from node $i$ to $j$. | The diameter of a SCN is the largest distance between any two firms in the network, in terms of number of intervening links on the shortest path. More complex manufacturing processes can include large network diameters (i.e. many stages of production) indicating difficulty in governing the overall SCN under a centralised authority. |
| **Network density (D)** | |
| $$D = \frac{<k>}{N-1}$$ where $<k>$ is the mean degree of all the nodes and $N$ is the total number of nodes, in the network | Density of a SCN indicates the level of interconnectivity between the firms involved. SCNs with high density indicate good levels of connectivity between firms which can be favourable in terms of efficient information exchange and improved robustness due to redundancy and flexibility (Sheffi and Rice, 2005). |
| **Network centralisation (C)** | |
| $$C = \frac{N}{N-2}\left(\frac{\max(k)}{N-1} - D\right)$$ where $N$ is the total number of nodes in the network and $\max(k)$ is the maximum degree of a node within the network. Density is determined as per the equation above. | Network centralisation provides a value for a given SCN between 0 (if all firms in the SCN have the same connectivity) and 1 (if the SCN has a star topology). This indicates how the operational authority is concentrated in a few central firms within the SCN. Highly centralised SCNs can have convenience in terms of centralised decision implementation and high level of controllability in production planning. However, highly centralised SCNs lack local responsiveness since relationships between firms in various tiers are decoupled (Kim et al., 2011). |
| **Network heterogeneity (H)** | |
| $$H = \frac{\sqrt{variance(k)}}{<k>}$$ where $<k>$ is the mean degree and $variance(k)$ is the variance | Heterogeneity is the coefficient of variation of the connectivity. Highly heterogeneous SCNs exhibit hub firms (i.e. firms with high number of contractual connections). In extreme cases, there may be many super large hubs (winner take all scenario, indicating centralised |
| of the degree, of all the nodes in the network. | control of the overall SCN through a single firm or a very few firms). |
| **Average clustering coefficient (<C>)** | |
| $$<C> = \frac{\sum_i C_i}{N}$$ where $N$ is the total number of nodes in the network and $C_i$ is the number of triangles connected to node $i$ divided by the number of triples centered around node $i$. | Clustering coefficient indicates the degree to which firms in a SCN tend to cluster together around a given firm. For example, it can indicate how various suppliers behave with respect to the final assembler at the global level (Kim et al., 2011). Therefore, the higher the clustering coefficient, the more dependent suppliers are on each other for production (Brintrup et al., 2016). |
| **Power-law exponent ($\gamma$) (Barabasi and Albert, 1999)** | |
| The degree distribution $P_k$ of a scale free network is approximated with power law as follows; $$P_k \sim k^{-\gamma}$$ where $k$ is the degree of the node and $\gamma$ is the power-law exponent (also known as the degree or scale free exponent). | SCNs with $\gamma<2$ include very large hubs which acquire control through contractual relationships with other firms at a rate faster than the growth of the SCN in terms of new firm additions. As $\gamma$ continues to increase beyond 2, the SCNs include smaller and less numerous hubs, which ultimately leads to a topology similar to that of a random network where all firms have almost the same number of connections.<br>Note that in directed networks, two $\gamma$ values are generally reported – one for the in-degree and another for the out-degree. |
| **Assortativity ($\rho$) (Newman, 2002)** | |
| Assortativity is defined as a correlation function of excess degree distributions and link distribution of a network.<br><br>For undirected networks, when degree distribution is denoted as $p_k$ and excess degree (remaining degree) distribution is denoted as $q_k$, one can introduce the quantity $e_{j,k}$ as the joint probability distribution of the remaining degree distribution of the remaining degrees of the two nodes at either end of a randomly chosen link.<br><br>Given these distributions, the assortativity of an undirected network is defined as;<br>$$\rho = \frac{1}{\sigma_q^2}\left[\sum_{jk} jk(e_{j,k} - q_j q_k)\right]$$ | Positive assortativity means that the firms with similar connectivity would have a higher tendency to connect with each other (for example, highly connected firms could be managing sub-communities in certain areas of production and then connect to other high-degree firms undertaking the same function). This structure can lead to cascading disruptions – where a disruption at one leaf node can spread quickly within the network through the connected hubs (Brintrup et al., 2016).<br><br>In contrast, a negative assortativity (i.e. disassortativity) indicates that it is the firms with dissimilar connectivity that tend to pair up in the given network. An unfavourable implication of disassortativity in SCNs is that since high degree firms are less connected to one another, many paths between nodes in the network are dependent on high degree nodes. Therefore, failure of a high degree node in a disassortative network would have a relatively large impact on the overall connectedness of the network (Noldus and Van, 2015).<br><br>On the other hand, disassortative networks are generally resilient against cascading impacts arising from targeted attacks – since hub nodes are not connected with each |
| where $\sigma_q$ is the standard deviation of $q_k$. | other, the likelihood of disruption impacts cascading from one hub node to another is minimised (Song et al, 2006). |
| **Modularity (Q) (Newman and Girvan, 2004)** ||
| $$Q = \sum_{s=1}^{k}\left[\frac{l_s}{L} - \left(\frac{d_s}{2L}\right)^2\right]$$ where $k$ is the number of modules, $L$ is the number of links in the network, $l_s$ is the number of links between nodes in module $s$, and $d_s$ is the sum of degrees of nodes in the module $s$. <br><br> To avoid getting a single module in all cases, this measure imposes $Q=0$ if all nodes are in the same module or nodes are placed randomly into modules. | SCNs with high modularity contain pronounced communities – i.e. partially segregated subsystems or modules embedded within the overall SCN system (Ravasz et al., 2002; Newman, 2003). Each of these subsystems are generally responsible for a particular specialised task. |
| **Percolation threshold for random node removal ($f_c$) (Cohen et al., 2000)** ||
| The percolation threshold for random node removal is given as; $$f_c = 1 - \frac{1}{\frac{<k^2>}{<k>} - 1}$$ where $<k>$ is the mean degree and $<k^2>$ is the second moment of the degree, of all the nodes in the network. | The percolation threshold of a SCN indicates the percentage of firms needed to be randomly removed prior to the overall SCN breaks into many disconnected components (when the giant component ceases to include all the nodes). In summary, this indicates the number of random firm failures that would drive the SCN from a connected state to a fragmented state (loss of overall interconnectivity). |

# APPENDIX B: LIST OF NODE LEVEL METRICS AND THEIR SCN IMPLICATIONS

**TABLE B: Node level metrics and their SCN implications**

| Mathematical representation | SCN Implication |
|---|---|
| **Degree ($k$)** | |
| The degree $k_i$ of any node $i$ is represented by; $$k_i = \sum_j a_{ij}$$ where $a_{ij}$ is any element of the adjacency matrix $A$. | Represents the number of direct neighbours (connections) a given firm has. For instance, in a given SCN, the firm with the highest degree (such as the integrators that assemble components) is deemed to have the largest impact on operational decisions and strategic behaviours of other firms in that particular SCN. Such a firm has the power to reconcile the differences between various other firms in the SCN and align their efforts with greater SCN goals (Kim et al., 2011).<br><br>In directed networks, the firms which have high in-degree are considered to be 'integrators' who collects information from various other firms to create high value products. In contrast, the firms which have high out-degree are considered to be 'allocators' who are generally responsible for distribution of high demand resources to other firms and/or customers. |
| **Betweenness centrality (normalised) (Freeman, 1977)** | |
| The betweenness centrality of a node $n$ is defined as; $$C_b(n) = \frac{2}{(N-1)(N-2)} \sum_{s \neq n \neq t} \frac{\sigma_{s,t}(n)}{\sigma_{s,t}}$$ where $s$ and $t$ are nodes in the network, which are different from $n$, $\sigma_{s,t}$ denotes the number of shortest paths from $s$ to $t$, and $\sigma_{s,t}(n)$ is the number of shortest paths from $s$ to $t$ that $n$ lies on. | Betweenness centrality of a firm is the number of shortest path relationships going through it, considering the shortest path relationships that connect any two given firms in the SCN. Therefore, it indicates the extent to which a firm can intervene over interactions among other firms in the SCN by being a gatekeeper for relationships. Those firms with high levels of betweenness generally play a vital role in SCNs – mainly owing to their ability to increase the overall efficiency of the SCN by smoothing various exchange processes between firms. |
| **Closeness centrality (Sabidussi, 1966)** | |
| The closeness centrality of a node n is defined as; $$C_c(n) = \frac{1}{<L(n,m)>}$$ where $<L(n,m)>$ is the length of the shortest path between two nodes $n$ and $m$ (note that for unweighted graphs with no geodesic distance | Closeness centrality is a measure of the time that it takes to spread the information from a particular firm to the other firms in the network. While it is closely related to betweenness centrality, closeness more relevant in situations where a firm acts as a generator of information (i.e. a navigator) rather than a mere mediator/gatekeeper.<br><br>For example, due to various hindrances, the market demand information can easily be distorted when it flows from the |
| information, each link is assumed to be one unit of distance). The closeness centrality of each node is a number between 0 and 1. | downstream firms towards upstream firms. Such distortions can lead to undue deviation between production plans of manufacturers and supply plans of suppliers, leading to a phenomenon known as the bullwhip effect in supply chains. Firms with high closeness centrality levels therefore play a major role in sharing the actual market demand information with upstream firms in the SCN, thus diminishing the adverse impacts arising from bullwhip effect (Xu et al., 2016). |
| **Eigenvector centrality (Ruhnau, 2000)** | |
| If the centrality scores of nodes are given by the matrix X and the adjacency matrix of the network is A, then each row of matrix X, namely x, can be defined as; $x \propto Ax$ i.e. $\lambda x = Ax$ The eigenvector centrality scores are obtained by solving this matrix equation. It can be shown that, while there can be many values for $\lambda$, only the largest value will result in positive scores for all nodes, so this is the eigenvector chosen. | Eigenvector centrality measures a firm's influence in the SCN by taking into account the influence of its neighbours. The centrality scores are given by the eigenvector associated with the largest eigenvalue. It assumes that the centrality score of a firm is proportional to the sum of the centrality scores of the neighbours. A firm with a high eigenvector centrality is assumed to derive its influential power through its highly connected neighbours. |

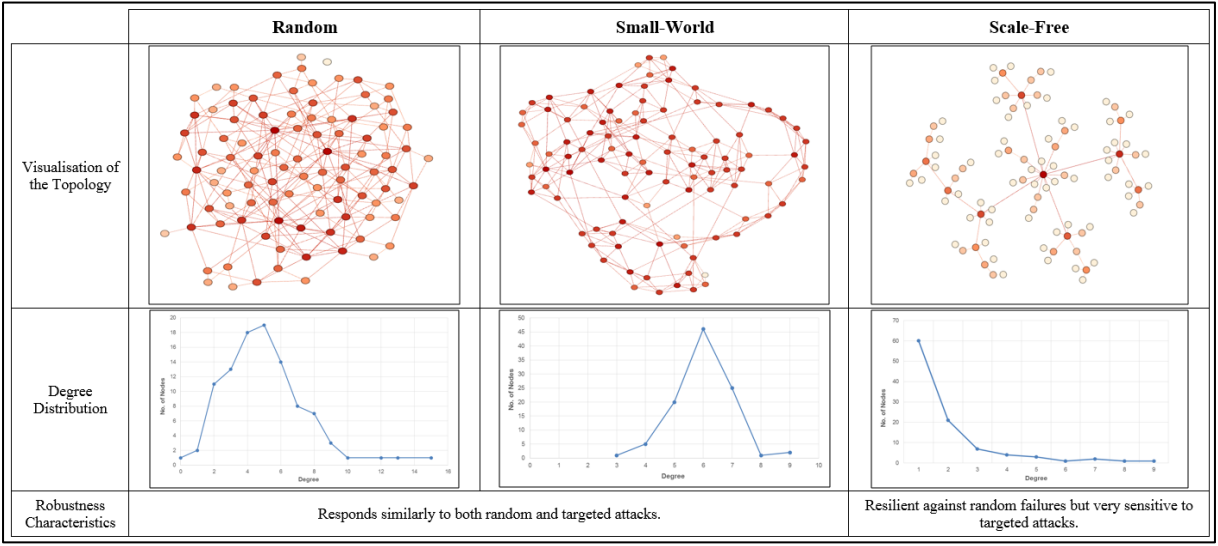

*Figure 1. Comparison of random, small-world and scale free networks*

*Topological structure of benchmark network models. Random and Small-world network topologies do not include hub nodes. In contrast, scale-free topologies are characterised by the presence of small number of highly connected hub nodes and a high number of feebly connected nodes. Presence of distinct hubs in scale-free networks make them more vulnerable to targeted attacks, compared to random and small-world networks.*

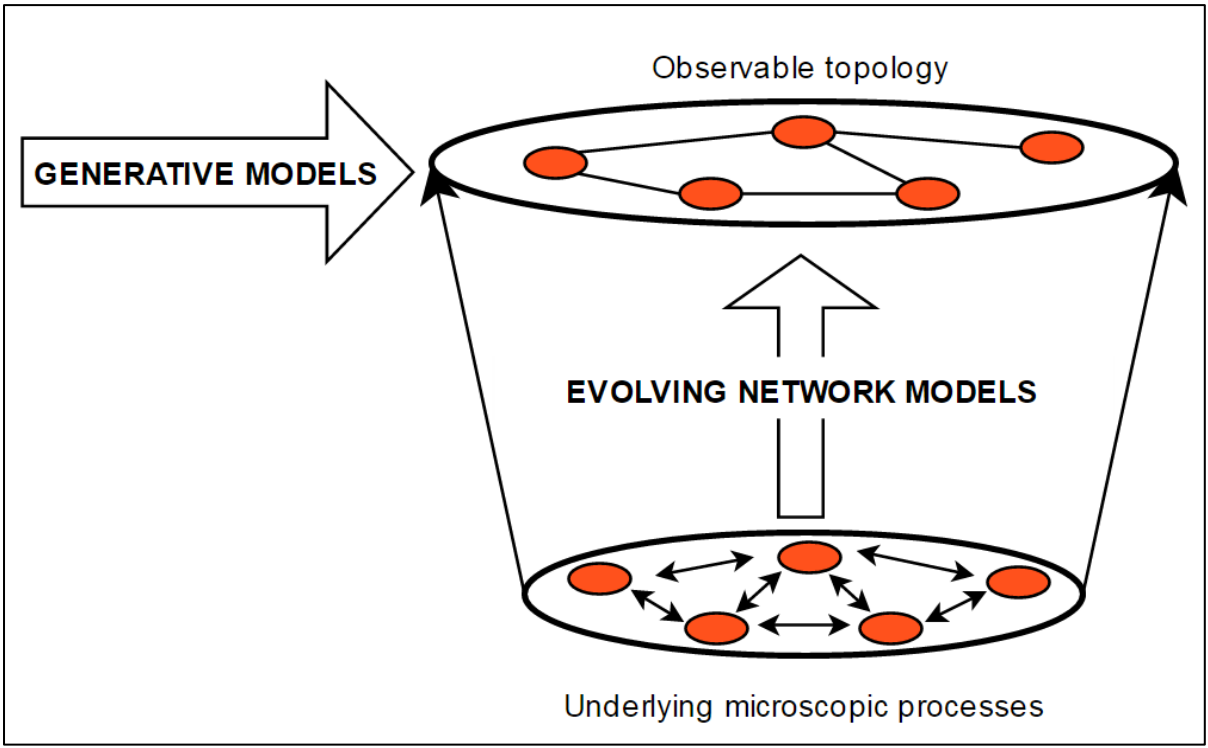

*Figure 2: Modelling Perspectives obtained from Generative and Evolving Network Models*

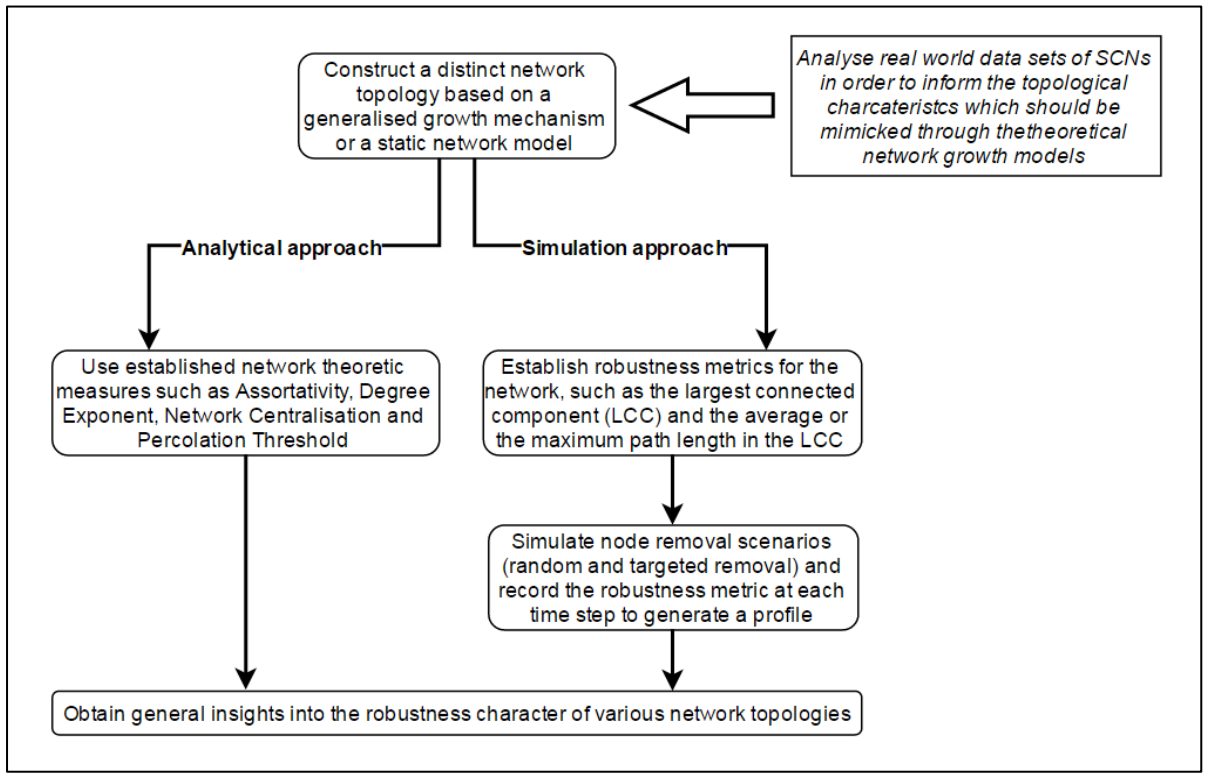

*Figure 3: General methodological framework of research on topology and robustness of SCNs*

| | | | | |
|---|---|---|---|---|
| **Shape parameter of the Lognormal distribution** | σ = 0.01 | σ = 2 | σ = 5 | σ = 10 |
| **Power law exponent** | γ > 3 | 3 > γ > 2 | γ < 2 | |
| **Description of topology** | Random | Scale-free | Hub and spoke | Winner-take-all |

*Figure 4: Transitions from random to winner-take-all graphs observed as σ parameter is increased*